\documentclass{mem}
\usepackage{natbib}\usepackage{txfonts}\usepackage{balance}
\usepackage{graphicx}
\usepackage[a4paper,breaklinks,dvipdfm]{hyperref}
\idline{75}{282}
\newcommand{\kms}{km\,s$^{-1}$}

\newcommand{\msun}{$M_{\odot}$}

\newcommand{\livii}{$^{7}$Li}

\begin{document}
\def\teff{$T\rm_{eff }$}
\def\kms{$\mathrm {km s}^{-1}$}

\title{
  Lithium in the closest satellite of our Milky Way
}

\author{
     G. \,Cescutti\inst{1,2} 
\and P. \,Molaro\inst{1,2}
\and X. \,Fu \inst{3}  }

 \institute{
   INAF, Osservatorio Astronomico di Trieste,
   Via G.B. Tiepolo 11, I-34143 Trieste, Italy
   \email{gabriele.cescutti@inaf.it}
\and
IFPU, Istitute for the Fundamental Physics of the
Universe, Via Beirut,  2, 34151, Grignano, Trieste, Italy
\and
The Kavli Institute for Astronomy and Astrophysics
at Peking University, Beijing 100871, China
}

\authorrunning{Cescutti et al.}

\titlerunning{Lithium in Gaia-Enceladus}

\abstract{
Recently, we studied the chemical evolution of lithium in the thin disc of the Milky Way. 
We found that the best agreement with the observed Li abundances in the thin disc is
 obtained considering novae as the main source of lithium. We assumed a delay time of $\approx$1 Gyr for nova
production and an effective \livii \, yield of 1.8($\pm$
0.6)x10$^{−5}$ \msun \, over the whole nova lifetime. 
The possibility to check our detailed assumptions on lithium production on other stellar systems, 
such as the satellites of our Milky Way, is seriously hampered by their distance from us. In these
 systems dwarf stars (where the original lithium can be measured) are too faint to detect lithium lines.
However, thanks to the Gaia mission, it was recently possible to disentangle the stars of a disrupted 
dwarf galaxy in the Galactic halo (called Enceladus or Galactic sausage). Adopting a chemical evolution
 model tuned to match the metallicity distribution function 
of Enceladus stars, we present our predictions for the lithium abundance of the stars of this disrupted galaxy. 

}

\maketitle{}

\section{Introduction}

In the last decades, kinematical and chemical surveys of the stars of
the Galactic halo revealed streams and structures belonging to
different stellar groups.
The Gaia survey has revealed a component in the inner halo showing a
peculiar velocity
and with metallicities $Z \approx Z_{\sun}/10$ which are relatively
more metal-rich than the Galactic halo
\citep{Helmi2018}. 
This structure called Gaia-Enceladus or also Gaia Sausage \citep{Belokurov2018}, likely
represents a disrupted dwarf galaxy after a major collision with the
Milky Way which happened more than 10 Gyr
ago. In the Galactic halo, a population of
stars showing low-[$\alpha$ /Fe] stellar component was already
identified and this was explained as stars possibly accreted from dwarf galaxies.
 Gaia-Enceladus  offers an unique opportunity to study the chemical
abundances of chemical elements in a dwarf galaxy which is normally
hampered by their large distance from us.  A detailed chemical
analysis of these stars was carried out in \citep{Vincenzo2019}. Here,
we focus on \livii\ in dwarf stars which are normally out of reach in
dwarf galaxies of the local group.

\section{Chemical evolution model for Gaia-Enceladus}\label{CEM}
\begin{figure}
   \includegraphics[width=\columnwidth]{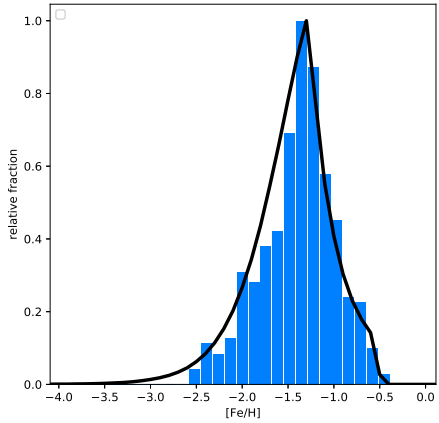}
\caption{Metallicity distribution function for Gaia-Enceladus. The
  histogram represents the observational data, the black line shows
  the model results.}
\label{MDF_ence}
\end{figure}

\begin{figure}
\includegraphics[width=\columnwidth]{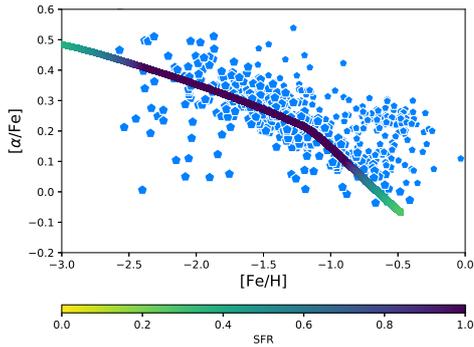}
\caption{[$\alpha$/Fe] as a function of [Fe/H]. The cyan pentagons are
  stars measured by APOGEE, which belong to Gaia-Enceladus according
  to \citet{Helmi2018}. The model is shown as a line and it is
  colour-coded according to the SFR (see colorbar).}
\label{alpha_ence}
\end{figure}

In this Section, we describe the main characteristics of a generic
chemical evolution models; moreover, we describe how we set its
parameters to match the particular chemical evolution of
Gaia-Enceladus; with this model we intend predict the Li abundance in
the Gaia-Enceladus.  The infall law is:
\begin{equation}
A(t)=M_{Enc} Gauss(\sigma_{Enc},tau_{Enc})
\end{equation}
where Gauss is a normalised Gaussian function, $\tau_{Enc}$ is
time of the center of the peak and $\sigma_{Enc}$ the standard
deviation; $M_{Enc}$ is the total amount of the gas accreted into
Gaia-Enceladus.  The star formation rate (SFR) is:
\begin{equation}
\psi(t)=
\bigg \{
\begin{array}{rl} 
\nu_{Enc} \Sigma(t)^k & t \leq T_{Enc} \\
0 & t > T_{Enc} \\
\end{array}
\end{equation}
where $\nu_{Enc}$ is the efficiency of the star formation, $\Sigma(r)$
is the surface mass density, and the exponent, $k$, is set equal to
1.5 \citep{Kennicutt89}; $T_{Enc}$ is the time when Gaia-Enceladus
stops forming star, due to the interaction with the Galaxy.  A
galactic wind is considered as follows:
\begin{equation}
W(t) =
\bigg \{
\begin{array}{rl} 
\nu^{wind}_{Enc}\psi(t) & t \geq T^{wind}_{Enc} \\
 0 & t < T^{wind}_{Enc} \\
\end{array}
\end{equation}
where $T^{wind}_{Enc}$ is when the galactic wind in Gaia-Enceladus
starts due to interaction with the Galaxy and $\nu^{wind}_{Enc}$ is
the wind efficiency.  Seven parameters - $\nu_{Enc}$,
M$_{Enc}$,$\tau_{Enc}$, $\sigma_{Enc}$ T$_{Enc}$, T$^{wind}_{Enc}$ and
$\nu^{wind}_{Enc}$ - determine the equations of the chemical evolution
model for Gaia-Enceladus.  These parameters are obtained by minimise
the results of the model and the chemical properties of the
kinematically selected stars of Gaia-Enceladus measured by the APOGEE
survey \citep{Helmi2018} and in particular the metallicity
distribution function (MDF); more details will be available in \citep{Cescutti2020} 
The best parameters are summarised in
Tab \ref{tab:tab_par}.
\begin{table}
 \caption{Best parameters for the chemical evolution of Gaia-Enceladus.}
 \label{tab:tab_par}
 \begin{tabular}{l|c}
  \hline
  parameter  & best value\\
  \hline
$\nu_{Enc}$ (star formation efficiency) & 1.3 Gyr $^{-1}$\\
M$_{Enc}$ (surface mass density) & 2.0 M$_{\odot}/pc^2$  \\
$\tau_{Enc}$ (peak of the infall law)& 550 Myr \\
$\sigma_{Enc}$ (SD of the infall law)& 1408 Myr \\
$\nu^{wind}_{Enc}$ (galactic wind efficiency) & 5.0 \\
T$^{wind}_{Enc}$ (start of the galactic wind) & 2919 Myr \\
T$_{Enc}$ (end of the star formation) & 5767 Myr  \\
 \hline
 \end{tabular}
\end{table}
The MDF obtained with the best parameters is compared to the
observational MDF of Gaia-Enceladus stars in Fig. \ref{MDF_ence}.
We obtain as best parameter for the surface
mass density of Gaia-Enceladus (M$_{Enc}$) 2 M$_{\odot}/pc^2$. This
value is $\approx$1/30 of the Galaxy, assuming as typical value the
solar vicinity with 50-60 M$_{\odot}/pc^2$. However, only 30\% of the
total mass is turned into stars during the evolution of
Gaia-Enceladus. Thus, a rough estimate of the total stellar mass of
Gaia-Enceladus should be approximately $<$10$^9$ M$_{\odot}$.  We also
compare the predicted [$\alpha$/Fe], with the abundances measured by
APOGEE. These data were not used to establish the best parameters. As
it is shown in Fig. \ref{alpha_ence}, the predictions of the model in
the [$\alpha$/Fe] vs [Fe/H] plane are in excellent agreement with the
data from the APOGEE survey.  
A minimise procedure was adopted also by \citet{Vincenzo2019} to
determine the chemical evolution model for Gaia-Enceladus. The main
difference is that \citet{Vincenzo2019} used the data in the
[$\alpha$/Fe] vs [Fe/H] to constrain the parameters of the chemical
evolution model, whereas in our approach the results of the model in
[$\alpha$/Fe] vs [Fe/H] are a prediction. Moreover, the galactic wind
in our Gaia-Enceladus model is due by the interaction with the Galaxy
and so $T^{wind}_{Enc}$ is a parameter. In \citet{Vincenzo2019}, it is
determined by number of supernovae explosions and the gravitational
potential of Gaia-Enceladus \citep{Lanfranchi06}.

\section{Lithium evolution model results in Gaia-Enceladus}
 \begin{figure}
\includegraphics[width=\columnwidth]{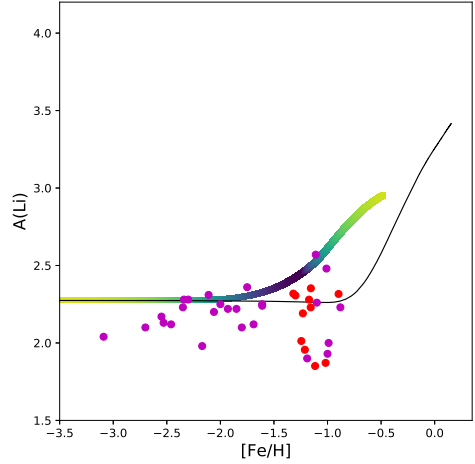}
\caption{Log(Li/H) vs [Fe/H]. Theoretical model results are
  presented with the observational data. The
  model is shown as a line colour-coded according to the SFR.
  For comparison we show with a black solid line
  the model results for the thin disc presented in
  \citet{Cescutti2019}. The observational data are shown as 
  solid circles. They are dwarf stars (logg$>$3.65) and
  Gaia-Enceladus members.  Red circles are stars from the
  GALAH survey, magenta dots from literature.}
\label{Li_ence}
\end{figure}
In Fig. \ref{Li_ence}, we show the predictions for \livii\ in
Gaia-Enceladus of our theoretical model, based on the assumptions on
the nucleosynthesis described in \citet{Cescutti2019} and the
parameters of chemical evolution for Gaia-Enceladus derived in
Sect.~\ref{CEM}. The sample of Gaia-Enceladus stars was obtained
by cross-matching the sample of 4644 confirmed Gaia-Enceladus
member stars \citep{Helmi2018} with the GALAH survey \citep{Buder18} and literature data \citep[for details,
see][]{Molaro20}.
The model predicts lithium abundances for
Gaia-Enceladus stars which rise from the Spite plateau at
[Fe/H]$\sim -$1.8.  On the other hand, the best model for the Milky Way
shows a similar increase, but at [Fe/H]$\sim -$1. Therefore, the model
predicts for the Gaia-Enceladus stars higher lithium abundances
compared to the genuine stars of the Milky Way of in the metallicity range $-$1
$<$[Fe/H]$<-$0.5. Clearly, this holds only for those stars that still
maintain their lithium original abundances (so dwarf stars with
T$>$5700K). The predicted overabundance is at maximum of 0.5 dex, but in most
cases is within the observation uncertainties. Still, it should be
feasible to assess this difference, and in the future with surveys
measuring a large number of stars in this metallicity region, a
statistical approach can be used to test our results.  At present, most
of the Gaia-Enceladus stars have similar behavior the one of the Milky Way
stars. Nevertheless, two stars present an enhancement of
lithium precisely on the model results predicted for Gaia-Enceladus.

\section{Discussion}
Recently several efforts have been made to measure {\it extragalactic}
\livii\ abundance.  Li measured belonging to Gaia-Enceladus are robust
"extragalactic " measurements of Li.  Omega Centauri ($\omega$ Cen) is
a globular cluster-like stellar system characterised by a wide range
of metallicities and probably ages. Usually, it is thought as the
stripped core of a dwarf galaxy. \citet{Monaco2010} found that dwarf
stars in $\omega$ Cen display a constant Li abundance and this is
observed among stars spanning a wide range of ages and metallicities
overlapping with the Spite plateau.  \citet{Mucciarelli2014} were able
to derive for the first time the initial lithium abundance in the
globular cluster M54 in the nucleus of the Sagittarius dwarf galaxy.
Sagittarius galaxy is located at 25 Kpc and its main sequence stars
are too faint ($\sim$22 mag) to be studied at high resolution. The
only possibility in this case are stars in the red giant branch
(RGB). However, the \livii\ abundance in RGB stars was modified by
stellar evolution and corrections are needed to account for the post
main sequence dilution.  By considering this dilution,
\citet{Mucciarelli2014} have established an initial Li abundance of
this stellar system of A(Li)= 2.29$\pm$ 0.11 (2.35$\pm$ 0.11
considering atomic diffusion, too).  The analysis of the
Gaia-Enceladus stars confirms the findings in Sagittarius and $\omega$
Cen: the Li problem seems to be a universal problem, regardless of
galaxy.  Therefore, the solution must work both in the Milky Way and
other galaxies, with different origins and star formation
histories. As noted by \citet{Mucciarelli2014}, it seems unlikely that
the scenario proposed by \citet{Piau2006}, requiring that at least one
third of the Galactic halo has been processed by Population III,
massive stars, can work in the same way in smaller systems like
Gaia-Enceladus.  According to our chemical evolution model results, we
expect that lithium will rise from the Spite plateau at a metallicity
lower in Gaia-Enceladus than in the thin disc of the Galaxy.  Future
observations aiming to relative hot dwarf stars of Gaia-Enceladus will
determine the correctness of this prediction.

\section{Acknowledgements}
G.C. acknowledges financial
support from the EU Horizon 2020 research and innovation
programme under the Marie Sklodowska-Curie grant agreement No. 664931
and from the EU COST Action CA16117 (ChETEC).
The GALAH survey is based on observations made at the Australian
Astronomical Observatory, under programmes A/2013B/13, A/2014A/25,
A/2015A/19, A/2017A/18. We acknowledge the traditional owners of the
land on which the AAT stands, the Gamilaraay people, and pay our
respects to elders past and present.



\bibliographystyle{aa}

\end{document}